\documentclass[10pt,fleqn]{article}

\usepackage{amsmath}
\usepackage{amssymb}

\usepackage{graphicx}

\usepackage{cite}

\usepackage{color}
\usepackage{comment}
\usepackage{longtable}
\usepackage{epstopdf}
\usepackage{morefloats}
\usepackage{ifetex}
\usepackage[colorlinks=true]{hyperref}
\hypersetup{urlcolor = blue, citecolor = red}
\usepackage[labelfont=bf,labelsep=period,justification=raggedright]{caption}

\makeatletter
\renewcommand{\@biblabel}[1]{\quad#1.}
\makeatother

\date{}

\begin{document}

\begin{flushleft}
{\Large
\textbf{Linking Cellular and Mechanical Processes in Articular Cartilage Lesion Formation: A Mathematical Model}
}
\\
Georgi I.~Kapitanov$^{1}$,
Xiayi Wang$^{2}$,
Bruce P.~Ayati$^{1,2,3}$,
Marc J.~Brouillette$^{3,4}$,
James A.~Martin$^{3,4}$
\\
\bf{1} Department of Mathematics, University of Iowa
\\
\bf{2} Program in Applied Mathematical \& Computational Sciences, University of Iowa
\\
\bf{3} Department of Orthopaedics \& Rehabilitation, University of Iowa
\\
\bf{4} Department of Biomedical Engineering, University of Iowa\\

\end{flushleft}

\section*{Abstract}
A severe application of stress on articular cartilage can initiate a cascade of biochemical reactions that can lead to the development of osteoarthritis. We constructed a multiscale mathematical model of the process with three components: cellular, chemical, and mechanical. The cellular component describes the different chondrocyte states according to the chemicals these cells release. The chemical component models the change in concentrations of those chemicals. The mechanical component contains a simulation of pressure application onto a cartilage explant and the resulting strains that initiate the biochemical processes. The model creates a framework for incorporating explicit mechanics, simulated by finite element analysis, into a theoretical biology framework.\\

Key words: articular cartilage, osteoarthritis, mathematical modeling, simulation, Abaqus, age-structured model, reaction-diffusion model


\section{Introduction}
\indent
The most common joint disease is osteoarthritis (OA), which causes joint pain and disability in those affected. It represents a growing cost to the healthcare system as the incidence is expected to increase from roughly 48 million people in 2005 to 65 million by 2030 \cite{Hootman:2006}. A subset of these cases develop after a known trauma and are then labeled post-traumatic OA (PTOA), and despite being heavily researched, the treatment options to prevent the occurrence of OA after an injury remain limited \cite{Anderson:2011}.
\

This is partly because the response of articular cartilage to compressive stress is complex. Moderate physiological stresses are known to be beneficial, causing the cells in cartilage, chondrocytes, to increase production of cartilage matrix molecules \cite{Sah:1989, Quinn:1998, Tomiyama:2007}. Severe applications of stress, such as a blunt impact injury or intra-articular fracture, cause chondrocyte death and eventual cartilage deterioration leading to PTOA development \cite{Anderson:2011}. However, the direct cell death from these impact injuries seems to be minor, as the majority of the cell death happens hours to days after the injury \cite{Martin:2009, Goodwin:2010, Tochigi:2011, Tochigi:2013}. A complex interplay between reactive oxygen species (ROS), pro-inflammatory cytokines (PIC), and erythropoietin (EPO) seems to dominate if/how this cell death occurs and spreads \cite{Martin:2009, Ding:2010, Goodwin:2010, Graham:2012, Wang:2015}. Creating a model that describes this pathway and determines thresholds that can cause the cell death to spread would provide an invaluable tool for identifying injuries that are at risk for PTOA lesion development or extrapolate treatment effects at varying doses or time scales that might not be feasible experimentally.
\

To that end, in the current article we present a multiscale mathematical model of the mechanotransductive processes that result from applying pressure with a metal indenter onto a cylindrical cartilage explant. The model consists of three scales: mechanical (tissue-level), cellular, and chemical. The mechanical component, external to the core biomathematical model and simulated with the finite element solver Abaqus\texttrademark , estimates the strains resulting from the pressure on the cartilage explant. The cellular component categorizes the behavior and states of chondrocytes under different chemical signals. The chemical component describes these chemical signals, as well as the cytokines the chondrocytes release.
\

Previous attempts to model the properties of cartilage only considered its mechanical behavior, for example \cite{Mow:1980, Mow:1989, Lai:1991}. We aim to build on the biomechanical approach by adding the interplay between the chondrocytes and cytokines, resulting in articular cartilage lesions.
 In previous work on this topic, the model in \cite{Wang:2014} considered constant cyclic loading on the cartilage, which has a different effect than the singular pressure that we are concerned with in the current study. In \cite{Graham:2012} and \cite{Wang:2014}, the authors used delay differential equations to describe the time delay in the switch of chondrocytes from one state to another. In \cite{Wang:2015}, the authors present an age-structured model that allows that switch to happen after a chondrocyte has been in a certain state for a certain amount of time. Age in this context is not actual cell age but how long a cell has been in its current state. This approach has led to faster computation times and greater model flexibility, and as such is used in the model in this paper. What is fundamentally different in the current model is that, unlike \cite{Wang:2015}, we integrate explicit mechanics into a biomathematical model in order to simulate the strain response of the tissue under pressure loading. This mechanical component uses explicit finite element analysis by computing the propagation of strain through the tissue from the applied pressure and calculates the starting density distribution of necrotic cells, which initiate the cellular and chemical components of the model.
 \

 This paper is organized as follows:  Section \ref{MM_section} introduces the model's equations and the numerical methods used for their solution. Section \ref{Results_section} describes the numerical results, and Section \ref{Discussion_section} is Discussion.

\section{Methods and Materials}\label{MM_section}
This section describes the mathematical model and the implemented numerical methods.

\subsection{Mathematical model}

The cartilage explant is modeled as a cylinder, with the assumption of circular symmetry. This assumption reduces the model to two dimensions in space: radial ($r$) and axial ($z$, representing the depth of the cylinder).  The independent variables of the system are radius ($r$), depth ($z$), time ($t$), and cell-state age ($a$).
\subsubsection{Components of the model}

\begin{figure}
\centering
\includegraphics[width=1\textwidth,]{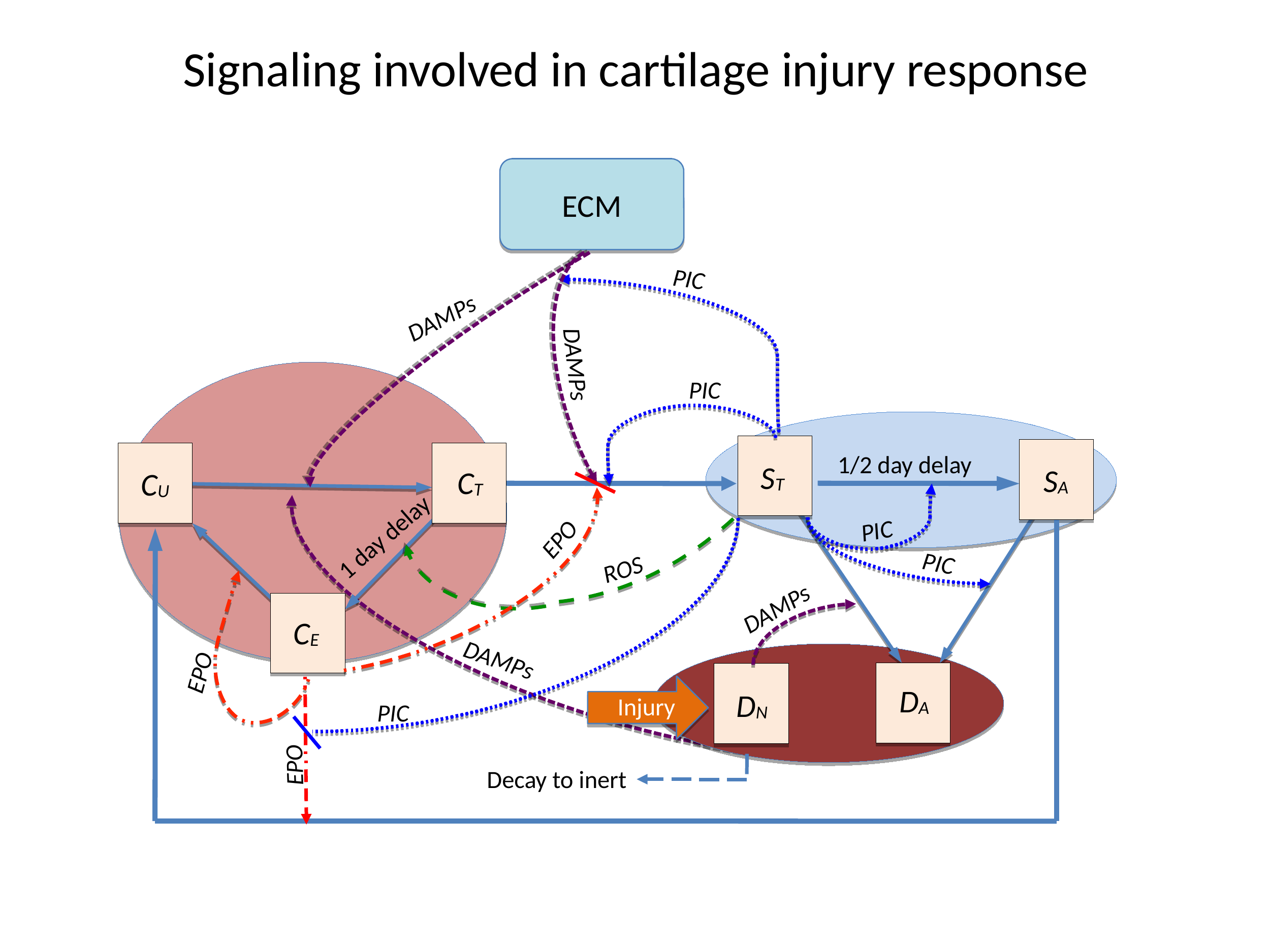}
\caption{Schematic of the articular cartilage lesion formation process due to one-time pressure load. The initial injury causes cell death. As a result the necrotic cells ($D_N$) release DAMPs, which initiate the chemical cascade leading to OA. Figure adapted from \protect\cite{Wang:2015}. Table \protect\ref{Var_table} contains a short description of the cell types and chemicals seen in the flowchart.}\label{FlowChart}  \end{figure}

A schematic of the system is presented in Figure \ref{FlowChart}. The pressure load causes necrosis of the chondrocytes. Necrotic chondrocytes release damage-associated molecular patterns (DAMPs), which cause the chondrocytes to release pro-inflammatory cytokines (PIC), such as tumor necrosis factor $\alpha$ (TNF-$\alpha$) and interleukin 6 (IL-6). Inflammation leads to cell apoptosis and the degradation of the extracellular matrix (ECM). However, it also signals the release of reactive oxygen species (ROS). ROS are precursors for the release of erythropoietin (EPO), which is an anti-inflammatory cytokine that counteracts the effects of inflammation. This complex feedback cycle requires two main components in our system of equations, cellular and chemical. The elements of the cellular component are:
\begin{itemize}
\item $C_U(r,z,t)$: population density (cells per unit area) of unsignaled healthy cells at a given time and location.
\item $C_T(r,z,a,t)$: population density of healthy cells signaled by DAMPs and in the process of becoming catabolic cells. In the presence of ROS they may be signaled to release EPO (by moving to the $C_E$ class) in 20-24 hours.
\item $C_E(r,z,t)$: population density of healthy cells signaled by ROS and starting to produce EPO.
\item $S_T(r,z,a,t)$: population density of cells in the catabolic state. Healthy cells signaled by alarmins (DAMPs) and PIC enter into the catabolic state. Catabolic cells synthesize cytokines associated with inflammation, and also produce ROS. $S_T$ cells can be signaled by the PIC to become EPOR-active cells ($S_A$). There is a 8-12-hour time gap before a cell expresses the EPO receptor after being signaled to become EPOR-active.
\item $S_A(r,z,t)$: population density of EPOR-active cells. EPOR-active cells express a receptor for EPO. Catabolic cells signaled by PIC enter the EPOR-active state and may switch back to healthy state $C_U$ when signaled by EPO.
\item $D_A(r,z,t)$: population density of  apoptotic cells. Catabolic cells are signaled by DAMPs and PIC to enter the apoptotic state. EPOR-active cells will also turn to apoptosis after signaled by PIC. Apoptotic cells are omitted in the system.
\item $D_N(r,z,t)$: population density of necrotic cells. In this model, necrotic cells emerge only due to the initial load. They release alarmins (DAMPs) into the system.
\end{itemize}
Chondrocytes exhibit minimal motility inside the ECM, so the cellular equations do not feature diffusion terms or other motility.
\

The elements of the chemical component are:
\begin{itemize}
\item $R(r,z,t)$: concentration of reactive oxygen species (ROS). In our model, ROS signal the pre-catabolic $C_T$ cells to start releasing EPO.
\item $M(r,z,t)$: concentration of alarmins (DAMPs) released by necrotic cells and ECM degradation. DAMPs signal healthy cells to enter a catabolic state, and together with pro-inflammatory cytokines, cause the catabolic cells $S_T$ to become apoptotic.
\item $F(r,z,t)$: concentration of general pro-inflammatory cytokines (PIC), e.g. TNF-$\alpha$ and IL-6, produced by catabolic cells ($S_T$). They have the following effects on the system:
  \begin{enumerate}
   \item[$-$] signal healthy cells ($C_T$) to enter the catabolic state ($S_T$),
   \item[$-$] signal catabolic cells ($S_T$) to enter the EPOR-active state,
   \item[$-$] cause both catabolic and EPOR-active cells to become apoptotic,
   \item[$-$] degrade the ECM, which in turn increases the level of DAMPs, resulting in further damage to the cartilage,
   \item[$-$] limit the production of EPO.
  \end{enumerate}
\item $P(r,z,t)$: concentration of erythropoietin (EPO), exclusively produced by $C_E$ cells in this model. Inflammation can suppresses this process. EPO helps EPOR-active cells ($S_A$) to switch back to the healthy state $C_U$. The effects of EPO depend on its concentration. When the concentration of EPO passes the threshold $P_c$ \cite{Brines:2008}, the spread of inflammation can be slowed by terminating the effect of the pro-inflammatory cytokines and DAMPs on the system. We also assume that $C_E$ cells revert to the $C_U$ state when the EPO level exceeds the $P_c$ threshold.
\end{itemize}
We assume that the chemicals diffuse through the entire region. The pro-inflammatory cytokines (PIC), such as TNF-$\alpha$ and IL-6, are the main promoter of cartilage lesion formation in this model, while EPO promotes cell recovery and limits the inflammation \cite{Brines:2008, Wojdasiewicz:2014, Eckardt:1989}. The balance between these pro-inflammatory and anti-inflammatory cytokines is essential for understanding the underlying causes of OA and is an important feature of the model.

\begin{itemize}
\item $U(r,z,t)$: density of the extracellular matrix (ECM). ECM is degraded by pro-inflammatory cytokines and releases DAMPs. The degradation of ECM is measured by the decrease in the concentration of SO$_4$ \cite{Farndale:1982}.\\
\end{itemize}
ECM degradation by proteases is simplified here to be related solely to pro-inflammatory cytokines, and expressed in terms of decreased suflate (SO$_4$) concentration. In cartilage, the proteoglycan groups, which comprise the majority of the ECM, contain sulfate groups and the measurement of sulfate concentration is related to ECM integrity or loss thereof. The average concentration of SO$_4$ in normal undamaged cartilage is $30$ g/L \cite{Farndale:1982}, which is the initial weight of ECM in this model.
Sufficiently high EPO concentration can also block ECM degradation.

\subsubsection{Equations}

The equations for the chemical concentrations are

\begin{align*}\label{ROS_eq}
        \tag{1a}
          \partial_t \underbrace{R(r,z,t)}_{\text{ROS}}=\underbrace{\frac{1}{r} \partial_{r}(r D_R  R_r) + \partial_{zz}D_RR}_{\text{diffusion}}-\underbrace{\delta_R R}_{\text{natural decay}}+\underbrace{\sigma_R S_T}_{\text{production by $S_T$}},
\end{align*}
\begin{align*}\label{DAMPs_eq}
        \tag{1b}
          \partial_t \underbrace{M(r,z,t)}_{\text{DAMPs}}=\underbrace{\frac{1}{r} \partial_{r}(r D_M  M_r)+ \partial_{zz}D_MM}_{\text{diffusion}}-\underbrace{\delta_M M}_{\text{natural decay}} +\underbrace{\sigma_M D_N}_{\text{production by $D_N$}} + \underbrace{\sigma_U U \frac{F}{\lambda_F +F}}_{\text{production by ECM}},
\end{align*}
\begin{align*}\label{PIC_eq}
      \tag{1c}
        \partial_t \underbrace{F(r,z,t)}_{\text{PIC}}=\underbrace{\frac{1}{r} \partial_{r}(r D_F  F_r)+ \partial_{zz}D_FF}_{\text{diffusion}}-\underbrace{\delta_F F}_{\text{natural decay}}+\underbrace{\sigma_F S_T}_{\text{production by $S_T$}},
\end{align*}
\begin{align*}\label{EPO_eq}
      \tag{1d}
      \partial_t \underbrace{P(r,z,t)}_{\text{EPO}}=\underbrace{\frac{1}{r} \partial_{r}(r D_P  P_r)+ \partial_{zz}D_PP}_{\text{diffusion}}-\underbrace{\delta_P P}_{\text{natural decay}} +\underbrace{\sigma_P C_E \frac{R}{\lambda_R+R} \frac{\Lambda}{\Lambda +F}}_{\text{production by $C_E$ controlled by PIC}},
\end{align*}
with no flux boundary conditions on the spatial domain
$0 \leq r \leq r_{max}$ and $0 \leq z \leq z_{max}$. The initial conditions are
\begin{align*}\label{Chem_ini_cond}
       \tag{1e}
       R(r,z,0)= M(r,z,0)=F(r,z,0)=P(r,z,0)=0.
\end{align*}

The Heaviside function used in several equations below is defined as
$$ H(\theta)=\begin{cases} 1, &  \theta \geq 0,\\
                                           0, & \theta < 0. \end{cases}$$
We use the Heaviside function to represent the cessation of inflammation when EPO exceeds a critical threshold ($P > P_c$).
\

The equation for the ECM density is

\begin{align*}\label{ECM_eq}
      \tag{2a}
        \partial_t \underbrace{U(r,z,t)}_{ECM}=\underbrace{-\delta_U U \frac{F}{\lambda_F +F} H(P_c - P)}_{\text{degradation by PIC under the control of EPO}},
\end{align*}
with initial condition
\begin{align*}\label{ECM_ini_cond}
      \tag{2b}
      U(r,z,0)=30 \text{mg}.
\end{align*}

The equations for the healthy cell population densities are
\begin{align*}\label{CU_eq}
        \tag{3a}
          \partial_t C_U(r,z,t)=\underbrace{\alpha_1 S_A \frac{P}{\lambda_P +P}}_{S_A \xrightarrow{\text{EPO}} C_U}+\underbrace{\alpha_2 H(P-P_c) C_E}_{C_E\xrightarrow{\text{EPO}} C_U}-\underbrace{\beta_{13} C_U\frac{M}{\lambda_M +M} }_{C_U\xrightarrow{\text{DAMPs}}C_T} ,
\end{align*}

\begin{align*}\label{CT_eq}
      \tag{3b}
       \partial_t C_T(r,z,a,t)+\partial_a C_T(r,z,a,t)= -\underbrace{\beta_{11}\frac{M}{\lambda_M +M}H(P_c - P)C_T(r,z,a,t)}_{C_T\xrightarrow{\text{DAMPs}}S_T} \\
                -\underbrace{\beta_{12} \frac{F}{\lambda_F +F} H(P_c-P)C_T(r,z,a,t)}_{C_T\xrightarrow{\text{PIC}}S_T}
                -\underbrace{\kappa_1 \gamma(a-\tau_2) \frac{R}{\lambda_R +R}C_T(r,z,a,t)}_{C_T\xrightarrow{\text{ROS}}C_E},
\end{align*}
\begin{align*}\label{CE_eq}
      \tag{3c}
       \partial_t C_E(r,z,t)= \underbrace{\int_0^{\infty} \kappa_1 \gamma(a-\tau_2) \frac{R(r,z,t)}{\lambda_R +R(r,z,t)}C_T(r,z,a,t) da}_{C_T\xrightarrow{\text{$\tau_2$ delay}}C_E}-\underbrace{\alpha_2 H(P-P_c) C_E}_{C_E\xrightarrow{\text{EPO}} C_U}.
\end{align*}

The function $\gamma(a)$ above represents the sharp age-dependent transition of cells from one state to another (in \eqref{CT_eq} the transition from $C_T$ to $C_E$ and later, in \eqref{ST_eq}, the transition from $S_T$ to $S_A$), in order to model the delay between the signal and the state-switch. It is given by:
\begin{equation}\label{gamma_def}
\gamma(a - a_{\text{max}}) = \frac{\gamma_0}{\sigma}\left(\tanh \left(\frac{a - a_{\text{max}}}{\sigma}\right) + 1\right),
\end{equation}
where $a_{\text{max}}$ is the state-age at which the cells switches states, $\gamma_0$ gives the height and $\sigma$ gives the spread of the function. The form of the function is taken from \cite{Wang:2015} and the $\gamma_0$ and $\sigma$ values are given in Table \ref{Table_Parameter}.
\

The tissue strains resulting from the pressure are only featured in the initial conditions, since there is no subsequent loading. Excessive strain causes necrosis in the initial population of healthy cells. The fraction of cells that die is dependent on the strain the cells withstand and is expressed through the function
\[\Gamma(\epsilon,r,z) = 0.01p_0(e^{K_U \epsilon} - e^{10 K_U}),\]
where $\epsilon$ is the absolute value of the position dependent axial (vertical) strain resulting from the deformation of the cartilage from the initial load, in \%. The constants $p_0$ and $K_U$ are parameter fitting constants. The form of the function is taken from \cite{Brouillette:2013}. Therefore, the initial number of healthy cells is
\begin{align*}\label{CU_ini_cond}
        \tag{3d}
        C_U(r,z,0)=(1-\Gamma(\epsilon,r,z))100,000 \text{ cells/cm}^2.
\end{align*}
The remaining initial and boundary conditions are
\begin{align*}\label{CT_bnd_cond}
      \tag{3e}
       C_T(r,z,0,t) =\underbrace{\beta_{13} C_U\frac{M}{\lambda_M +M} }_{C_U\xrightarrow{\text{DAMPs}}C_T} ,
\end{align*}
\begin{align*}\label{CT_ini_cond}
      \tag{3f}
       C_T(r,z,a,0)=C_E(r,z,0)=0.
\end{align*}

The equations for the sick cell population densities are
\begin{align*}\label{ST_eq}
      \tag{4a}
       \partial_t S_T(r,z,a,t)+\partial_a S_T(r,z,a,t)= &-\underbrace{\mu_{ST} \frac{F}{\lambda_F +F}\frac{M}{\lambda_M +M}S_T(r,z,a,t)}_{S_T\xrightarrow{\text{PIC, DAMPs}}D_A}\\
        &-\underbrace{\kappa_2 \cdot \gamma(a-\tau_1) \frac{F}{\lambda_F +F}S_T(r,z,a,t)}_{S_T\xrightarrow{\text{PIC}} S_A},
\end{align*}
\begin{align*}\label{SA_eq}
      \tag{4b}
       \partial_t S_A(r,z,t)=&\underbrace{\int_0^{\infty} \kappa_2 \cdot \gamma(a-\tau_1) \frac{F}{\lambda_F +F}H(P_c-P)S_T(r,z,a,t) da}_{S_T\xrightarrow{\text{$\tau_1$ delay}} S_A} -\underbrace{\alpha_1 S_A \frac{P}{\lambda_P +P}}_{S_A \xrightarrow{\text{EPO}} C_U} \\
       &-\underbrace{\mu_{S_A} \frac{F}{\lambda_F +F}H(P_c-P)S_A}_{S_A \xrightarrow{\text{PIC}} D_A} ,
\end{align*}
with initial and boundary conditions
\begin{align*}\label{ST_bnd_cond}
      \tag{4c}
       S_T(r,z,0,t) =\int_0^{\infty}(\underbrace{\beta_{11} \frac{M}{\lambda_M +M} H(P_c - P)}_{C_T\xrightarrow{\text{DAMPs}}S_T} +\underbrace{\beta_{12} \frac{F}{\lambda_F +F} H(P_c - P)}_{C_T\xrightarrow{\text{PIC}}S_T} ) C_T(r,z,a,t) da,
\end{align*}
\begin{align*}\label{Sick_ini_cond}
      \tag{4d}
       S_T(r,z,a,0)=S_A(r,z,0)=0.
\end{align*}

We track necrotic cells using
\begin{align*}\label{DN_eq}
      \tag{5a}
       \partial_t D_N(r,z,t)= -\underbrace{\mu_{D_N} D_N}_{\text{natural decay}},
\end{align*}
with initial condition
\begin{align*}\label{DN_ini_cond}
      \tag{5b}
D_N(r,z,0)= \Gamma(\epsilon,r,z)100,000 \text{ cells/cm}^2.
\end{align*}
There is no equation for apoptotic cells; these are considered removed from the system.\\
Short descriptions of all variables are in Table \ref{Var_table}

\begin{table}[ht]
\centering
\begin{tabular}{|l|l|}
\hline
 Variable & Description  \\
 \hline
 $C_U$& Healthy unsignaled chondrocytes.\\
 \hline
 $C_T$& Chondrocytes signaled by DAMPs to become catabolic.\\
 \hline
 $C_E$&  Chondrocytes that produce EPO.\\
 \hline
 $S_T$ & Catabolic cells that release PIC and ROS.\\
 \hline
 $S_A$ & EPOR-active cells, can switch back to $C_U$ if signaled by EPO.\\
 \hline
 $D_N$ & Necrotic cells. Formed by initial strain, release DAMPs.\\
 \hline
 $D_A$ & Apoptotic cells, not included in the model.\\
 \hline
 $R$ (ROS) & Reactive oxygen species. Signal $C_E$ cells to release EPO.\\
 \hline
 $M$ (DAMPs) & Damage-associated molecular patterns. Alarmins released by $D_N$ and the ECM as a\\
 &result of the loading.\\
 \hline
 $F$ (PIC) & Pro-inflammatory cytokines. Signal healthy cells to become catabolic and catabolic cells\\
 &to become apoptotic. Degrade the ECM.\\
 \hline
 $P$ (EPO) & Erythropoietin. Anti-inflammatory cytokine. Diminishes the effects of inflammation. \\
 \hline
 $U$ (ECM) & Extracellular matrix. Chondrocytes live within it. Subject to degradation from\\
 &the stress application and PIC.\\
 \hline
\end{tabular}
\caption{Cell types and variable meaning.}
\label{Var_table}
\end{table}

\subsubsection{Numerical Implementation}
Our computations are done in two main stages: a finite element analysis of the strain due to the initial load using the commercial software Abaqus\texttrademark , followed by a simulation of the cellular and biochemical response of the explant using our own software.
\

Abaqus\texttrademark is a finite element analysis (FEA) software developed by Dassault Syst\`emes. It is generally used in a variety of engineering projects, particularly for predicting mechanical stresses and strains on automotive designs under static and dynamics loads. Because of its accuracy and wide material modeling capability it has become one of the most widely used finite element solvers in biomedical fields.
\

Our in-house software features a step-doubling alternating-direction implicit (ADI) method, described in mathematical detail in \cite{Ayati:2005, Ayati:2006}, for the time and 2D space integration. The age discretization is done through a ``natural-grid'' Galerkin method, which to our knowledge remains the state of the art for solving partial differential equations that depend on age as well as time and space.   The mathematical underpinnings and general description of the natural-grid methods are in \cite{Ayati:2000, Ayati:2002, Ayati:2007, Ayati:2009}. The natural-grid approach has been used for the modeling and simulation of a range of systems, such as {\em Proteus mirabilis} swarm colony development \cite{Ayati:2006_swarm, Ayati:2007_swarm, Ayati:2009_swarm}, avascular tumor invasion \cite{Ayati:2006}, biofilm persistence and senescence \cite{Ayati:2007-Klapper, Klapper:2007}, bacterial dormancy \cite{Ayati:2012_JTB, Ayati:2012}, and now articular cartilage lesion formation \cite{Wang:2015}.

\subsubsection{Mechanical Component and Strain Simulation}
\begin{figure}
\centering
\includegraphics[width=0.7\textwidth, scale = 2]{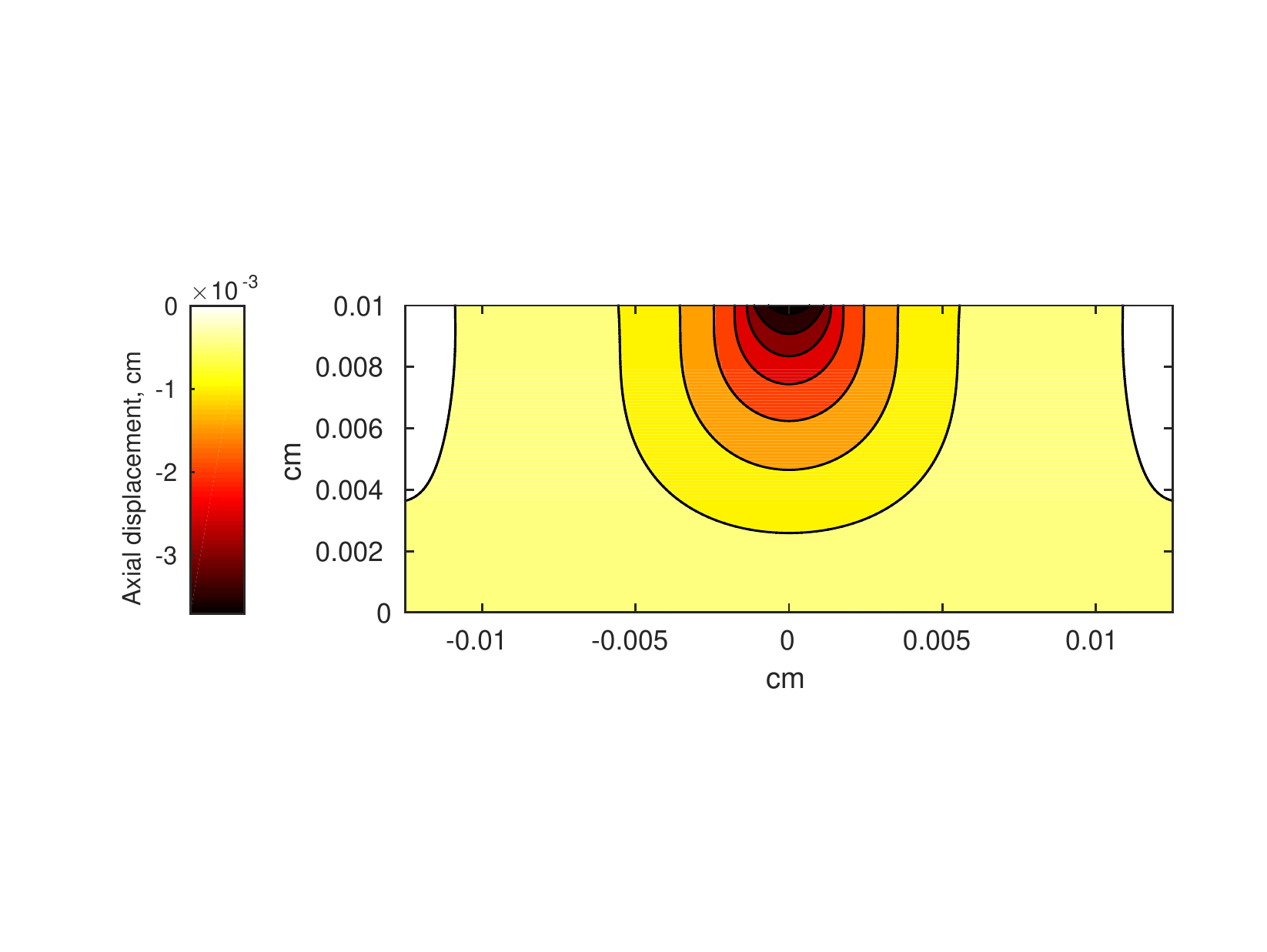}
\caption{Axial displacement on the cartilage rectangle, plotted with MATLAB, using raw data from Abaqus\texttrademark .}  \label{ABQ_displ} \end{figure}

\begin{figure}
\centering
\includegraphics[width=0.7\textwidth, scale = 2]{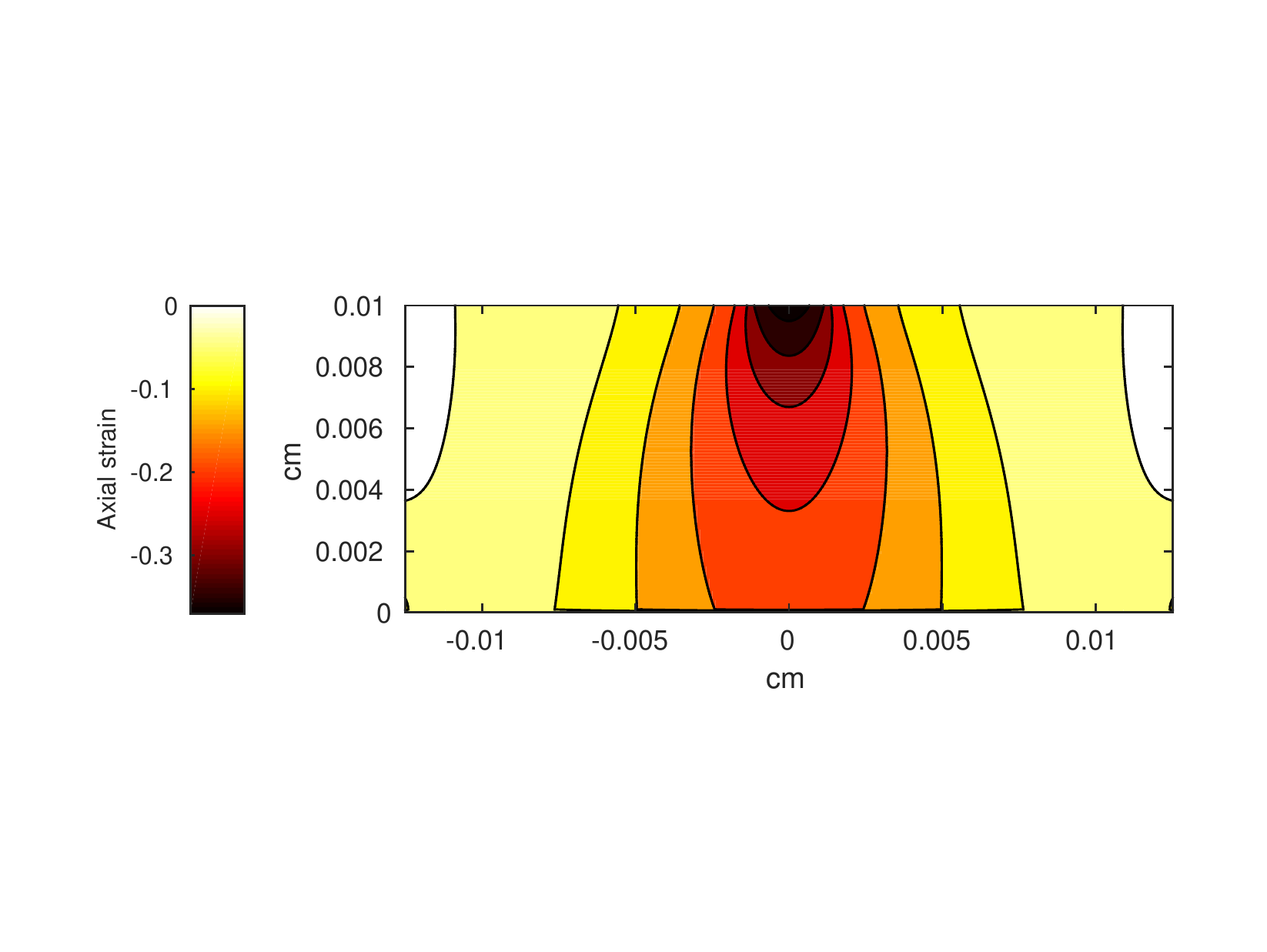}
\caption{Axial strain on the cartilage rectangle, calculated from raw displacement data from Abaqus\texttrademark .}  \label{ABQ_strain} \end{figure}

The mechanical component of the modeling and simulation involves the calculations of different strains across the spatial domain of the cartilage explant, and is used in the initial conditions \eqref{CU_ini_cond} and \eqref{DN_ini_cond} above. For the purposes of the current model, cartilage is considered to be homogeneous tissue, although in reality it is composed of several layers with different mechanical stiffness. Another assumption is that the cartilage disk exhibits linearly elastic behavior. Linear elasticity is generally modeled using hyperbolic partial differential equations. Since the external pressure only affects the initial conditions, there is no feedback between our parabolic system and the strains resulting from the initial load. Therefore, the strain at each points is constant relative to the system. The commercial finite element solver Abaqus\texttrademark was used to simulate a static pressure onto a rectangular block and record the resulting displacements over the spatial domain, from which we then calculated the axial strains. The axial strains were calculated by computing the ratio of vertical displacement and the vertical position of the node at which the displacement was measured.
 The axial strains depend on the pressure, as well as the radial and depth positions. The Abaqus\texttrademark simulation was done using a rectangle of dimensions $2.5\ \text{cm} \times 1.0\ \text{cm}$. A pressure of 0.4 MPa was applied onto a line of length 5.5 mm (to simulate an indenter) in the center top side of the rectangle. Since our model is two-dimensional in space, a two-dimensional Abaqus\texttrademark simulation is sufficient. In order to simulate the pressure applied by the indenter, the load is of the instantaneous pressure type. The Abaqus\texttrademark output is the axial displacement, $U_{2}$, at each point of the rectangle. The mesh contains over 250000 grid points. Because of the symmetry of the results, we only recorded the right half of the rectangle's strains and used those data in the initial conditions of our model. The Abaqus\texttrademark displacement output can be seen in Figure \ref{ABQ_displ}. The calculated strains are in Figure \ref{ABQ_strain}.

\section{Results}\label{Results_section}
\subsection{Parameter estimation}
 All parameters and a reference for their value are listed in Table \ref{Table_Parameter}. A detailed reasoning for selecting some of the parameters based on the literature is presented in \cite{Wang:2014}. The strain distribution across the simulated cylinder was computed by Abaqus\texttrademark and required several parameters related to the physical properties of cartilage: Young's modulus, Poisson ratio, and density. Since cartilage has a heterogeneous structure, these parameters can be estimated within a range. The values we used for the simulation were chosen as a combination of known values taken from \cite{Mansour:2003} and values within that range that corresponded to the findings related to pressure, strain, and cell death in \cite{Brouillette:2013}. The Young's modulus, Poisson ration, and density used in the computations are also in Table \ref{Table_Parameter}.
 \begin{table}[!ht]
\centering
    \begin{tabular}{|c|c|c|c|}
         \hline
           Parameter &  Value & Units & Reason\\
         \hline
           $D_R$ & 0.1 & $\frac{\text{cm}^2}{\text{day}}$ & Determined in \cite{Graham:2012}\\
         \hline
           $D_M$ & 0.05 & $\frac{\text{cm}^2}{\text{day}}$ & Determined in \cite{Graham:2012}\\
         \hline
           $D_P$ & 0.005 & $\frac{\text{cm}^2}{\text{day}}$ & Determined in \cite{Graham:2012}\\
         \hline
           $D_F$ & 0.05 & $\frac{\text{cm}^2}{\text{day}}$ & Determined in \cite{Graham:2012}\\
         \hline
           $\delta_R$ & 60& $\frac{1}{\text{day}}$ & Determined in \cite{Wang:2014}\\
         \hline
           $\delta_M$ & 0.5545 & $\frac{1}{\text{day}}$ & Determined in \cite{Wang:2014}\\
         \hline
           $\delta_F$ & 0.1664 & $\frac{1}{\text{day}}$ & Determined in \cite{Wang:2014}\\
         \hline
           $\delta_P$ & 3.326 & $\frac{1}{\text{day}}$ & Determined in \cite{Wang:2014}\\
         \hline
           $\delta_U$ & 0.0193& $\frac{1}{\text{day}}$ & Determined in \cite{Wang:2014}\\
         \hline
           $\sigma_R$ & 0.0024 & $\frac{\text{nanomolar}\cdot \text{cm}^2}{\text{day}\cdot \text{cells}}$ & Determined in \cite{Wang:2014}\\
         \hline
           $\sigma_M$ & 5.17$\times 10^{-7}$& $\frac{\text{nanomolar}\cdot \text{cm}^2}{\text{day}\cdot \text{cells}}$ & Determined in \cite{Wang:2014}\\
         \hline
           $\sigma_F$ & 2.35$\times 10^{-7}$ & $\frac{\text{nanomolar}\cdot \text{cm}^2}{\text{day}\cdot \text{cells}}$ & Determined in \cite{Wang:2014}\\
         \hline
           $\sigma_P$ &4.2$\times 10^{-5}$ & $\frac{\text{nanomolar}\cdot \text{cm}^2}{\text{day}\cdot \text{cells}}$ & Determined in \cite{Wang:2014}\\
         \hline
           $\sigma_U$ & 0.0154 & $\frac{\text{nanomolar}\cdot \text{cm}^2}{\text{day}\cdot \text{cells}}$ & Determined in \cite{Wang:2014}\\
         \hline
           $\Lambda$ & 0.5& nanomolar & Estimated\\
         \hline
           $\lambda_R$ & 5 & nanomolar  & Estimated\\
         \hline
           $\lambda_M$ & 0.5 & nanomolar  & Estimated\\
         \hline
           $\lambda_F$ & 0.5 & nanomolar  & Estimated\\
         \hline
           $\lambda_P$ & 0.5 & nanomolar & Estimated\\
         \hline
           $\alpha_1$ & 1 & $\frac{1}{\text{day}}$ & Estimated\\
         \hline
           $\alpha_2$ & 1 & $\frac{1}{\text{day}}$ & Estimated\\
         \hline
           $\beta_{11}$ & 100 & $\frac{1}{\text{day}}$ &Estimated\\
         \hline
           $\beta_{12}$ & 50 & $\frac{1}{\text{day}}$ & Estimated\\
         \hline
           $\beta_{13}$ & 10 & $\frac{1}{\text{day}}$ & Estimated\\
         \hline
           $\kappa_1$ & 10 & $\frac{1}{\text{day}}$ & Estimated\\
         \hline
           $\kappa_2$ & 10 & $\frac{1}{\text{day}}$ & Estimated\\
         \hline
           $P_c$ & 1 & nanomolar & Determined in \cite{Wang:2014} \\
          \hline
           $\mu_{S_T}$ & 0.5 & $\frac{1}{\text{day}}$ & Estimated\\
         \hline
           $\mu_{S_A}$ & 0.1 & $\frac{1}{\text{day}}$ &Estimated\\
         \hline
           $\mu_{D_N}$ & 0.05& $\frac{1}{\text{day}}$ &Estimated\\
         \hline
           $\tau_1$ & 0.5 & days &Determined in \cite{Graham:2012}\\
         \hline
           $\tau_2$ & 1 & days &Determined in \cite{Graham:2012}\\
           \hline
           $\gamma_0$ & 1 & & Determined in \cite{Wang:2015}\\
           \hline
           $\sigma$ & 0.1 & & Determined in \cite{Wang:2015}\\
           \hline
           $p_0$  & 1  & & Determined in \cite{Brouillette:2013}\\
           \hline
           $K_U$ & 0.0545 & & Determined in \cite{Brouillette:2013}\\
           \hline
           Abaqus\texttrademark Cartilage density & 1600 & kg/m$^3$ & Taken from \cite{Mansour:2003}\\
           \hline
           Abaqus\texttrademark Cartilage Young's modulus & $5 \times 10^5$& $Pa$ & Taken from \cite{Mansour:2003}\\
           \hline
           Abaqus\texttrademark Cartilage Poisson ratio & 0.25 & &Taken from \cite{Mansour:2003}\\
         \hline
    \end{tabular}
    \caption{ Table of parameters}
\label{Table_Parameter}
 \end{table}
\subsection{Computational results and experimental validation}
\begin{figure}
\centering
\includegraphics[width=0.7\textwidth]{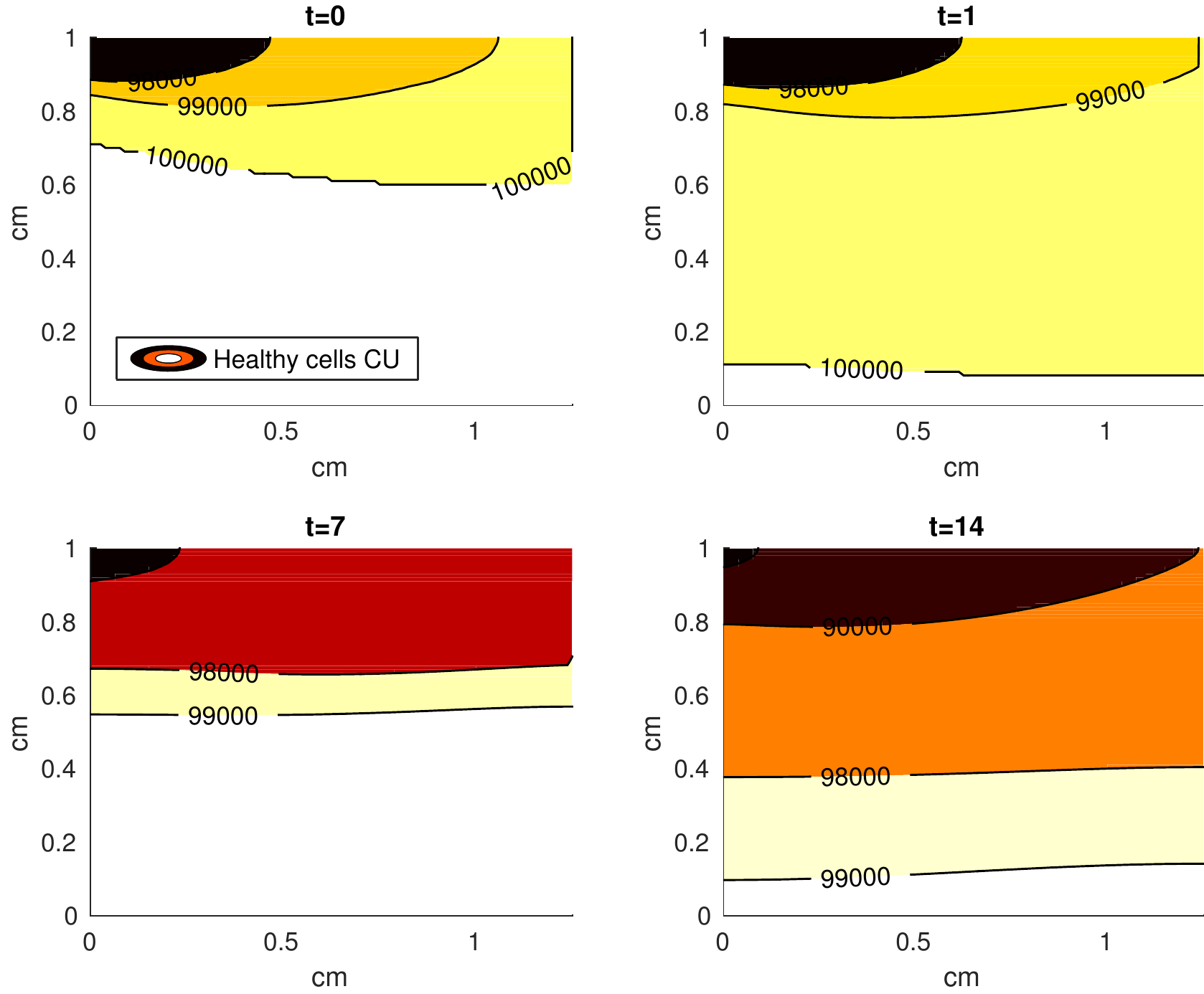}
\caption{A contour graph of the density of the $C_U$ cells in the 2D model at 0, 1, 7, and 14 days. The contour plot shows the decrease of $C_U$ cell density in the rectangular representation of the cylindrical explant. The lowest number of cells is in the upper left corner, the area of the initial contact.}\label{CU_contour}  \end{figure}

\begin{figure}
\centering
\includegraphics[width=0.7\textwidth]{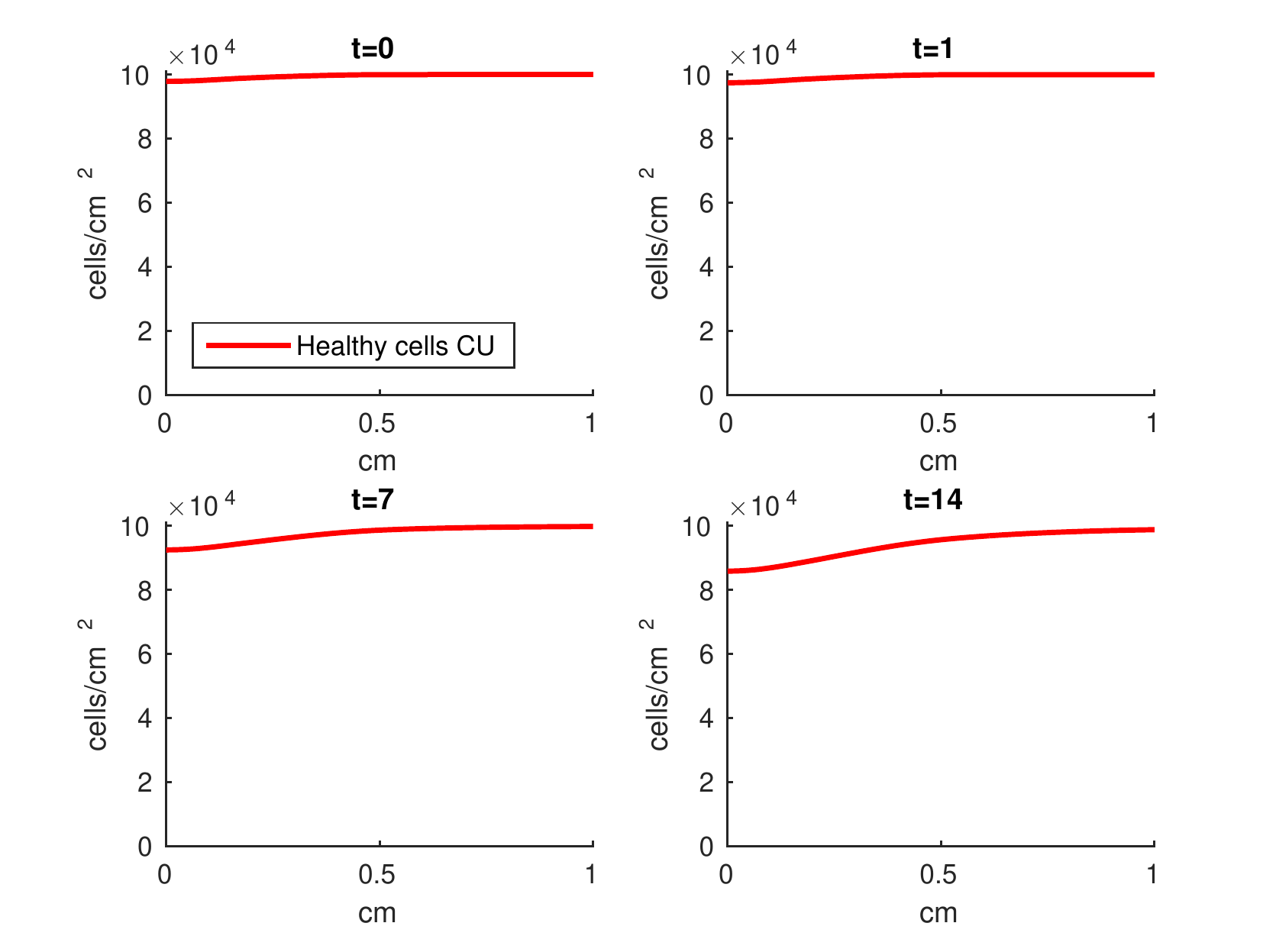}
\caption{Average density of $C_U$ cells along the radius of the model cylinder at 0, 1, 7, and 14 days. The $C_U$ cells are less dense near the area of contact.} \label{CU_ave} \end{figure}

\begin{figure}
\centering
\includegraphics[width=0.7\textwidth]{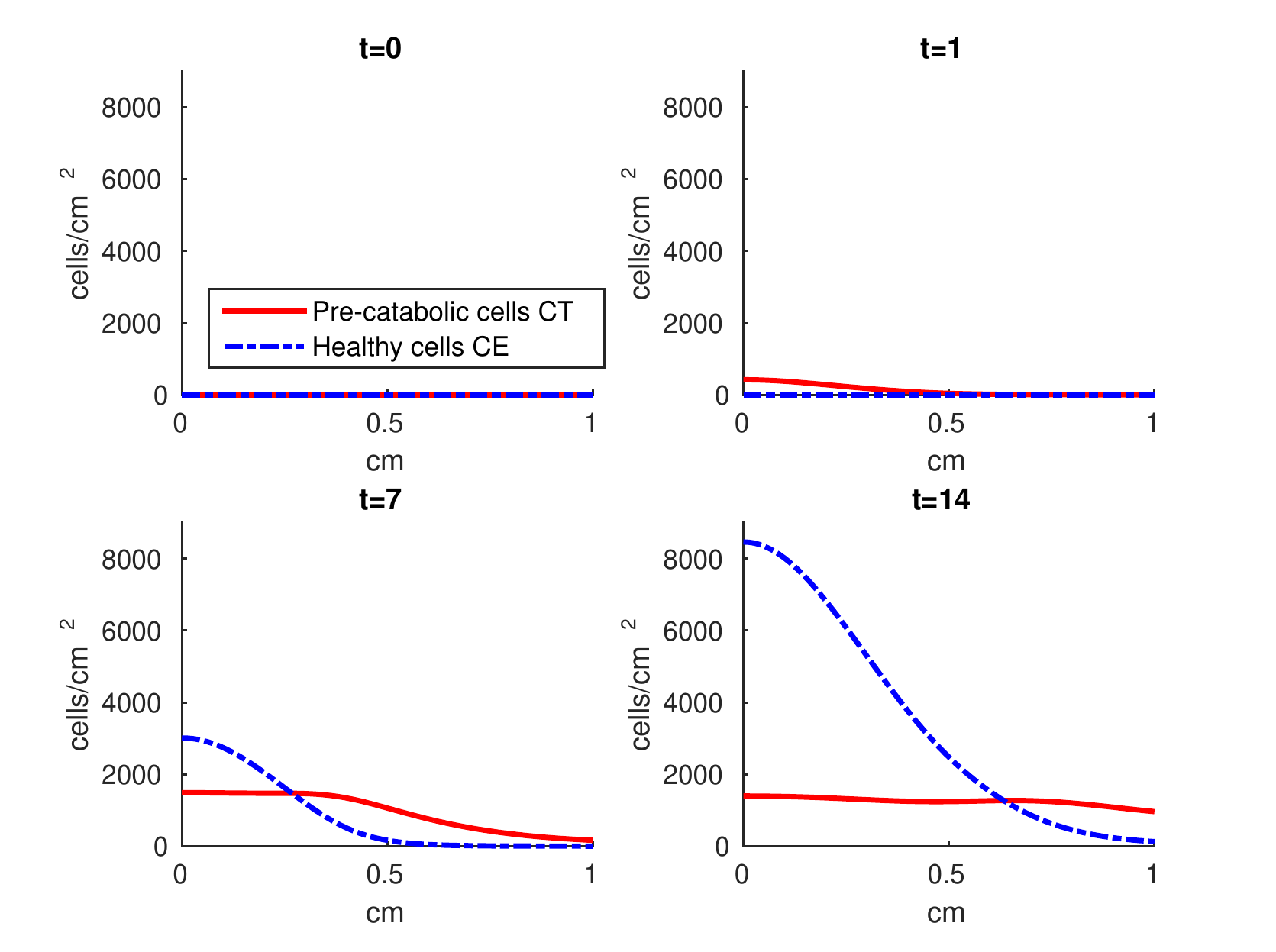}
\caption{Average density of $C_T$ and $C_E$ cells along the radius of the model cylinder at 0, 1, 7, and 14 days. The density of $C_E$ cells is higher near the contact area, while the density of the $C_T$ cells is more evenly distributed.}\label{CT_CE_ave}  \end{figure}

\begin{figure}
\centering
\includegraphics[width=0.7\textwidth]{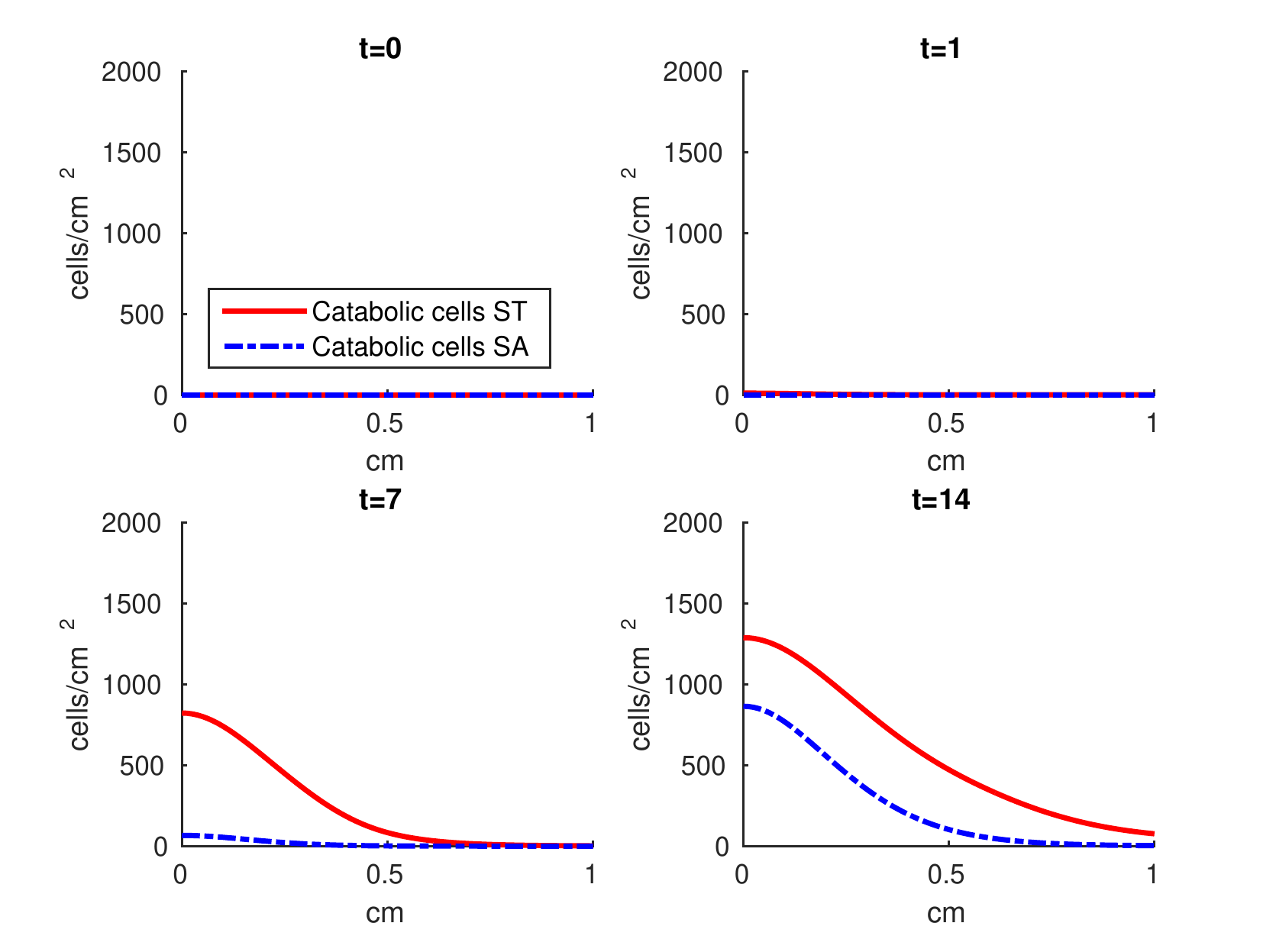}
\caption{Average density of $S_T$ and $S_A$ cells along the radius of the model cylinder at 0, 1, 7, and 14 days. The densities of both cell states are higher near the contact area.}\label{ST_SA_ave}  \end{figure}

\begin{figure}
\centering
\includegraphics[width=0.7\textwidth]{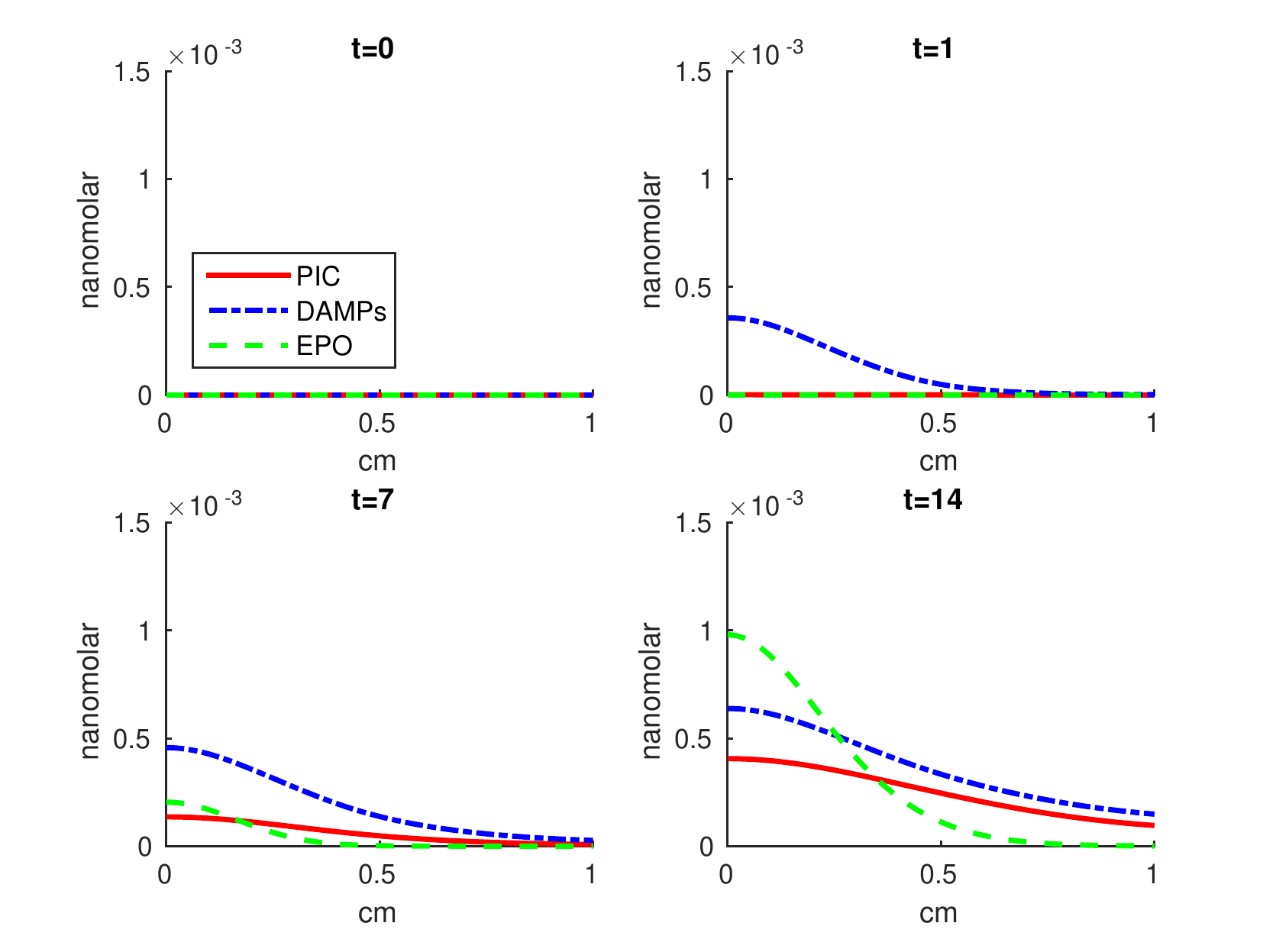}
\caption{Average concentrations of EPO, DAMPs, and PIC along the radius of the model cylinder at 0, 1, 7, and 14 days. DAMPs and particularly EPO concentrations are higher near the contact area. PIC concentrations are more evenly spread, although their concentrations are also somewhat higher near the area of contact.}\label{Chem_ave}  \end{figure}

\begin{figure}
\centering
\includegraphics[width=0.7\textwidth]{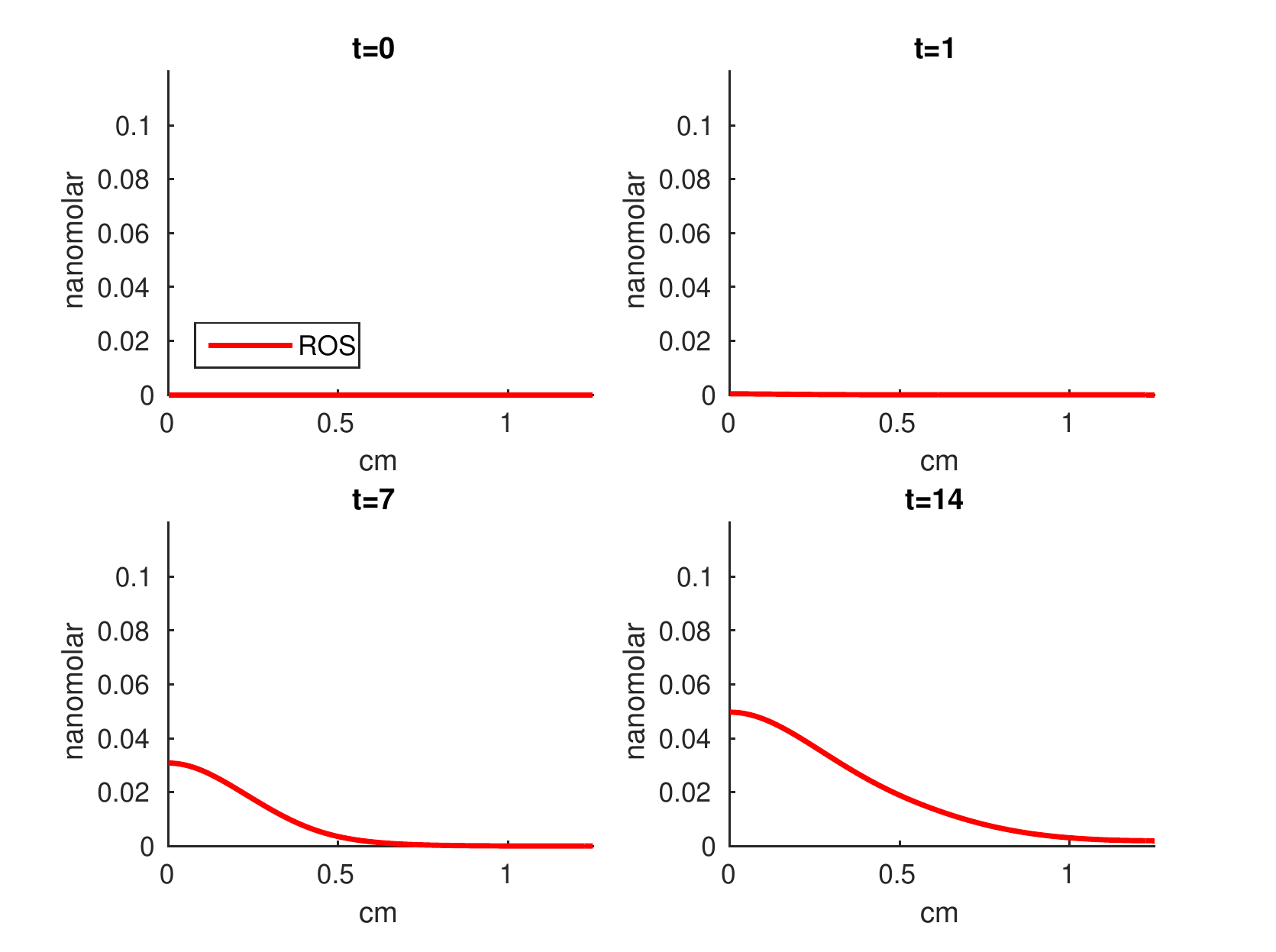}
\caption{Average concentrations of ROS along the radius of the model cylinder at 0, 1, 7, and 14 days. The concentrations are higher near the area of contact.}\label{ROS_ave}  \end{figure}

\begin{figure}
\centering
\includegraphics[width=0.7\textwidth]{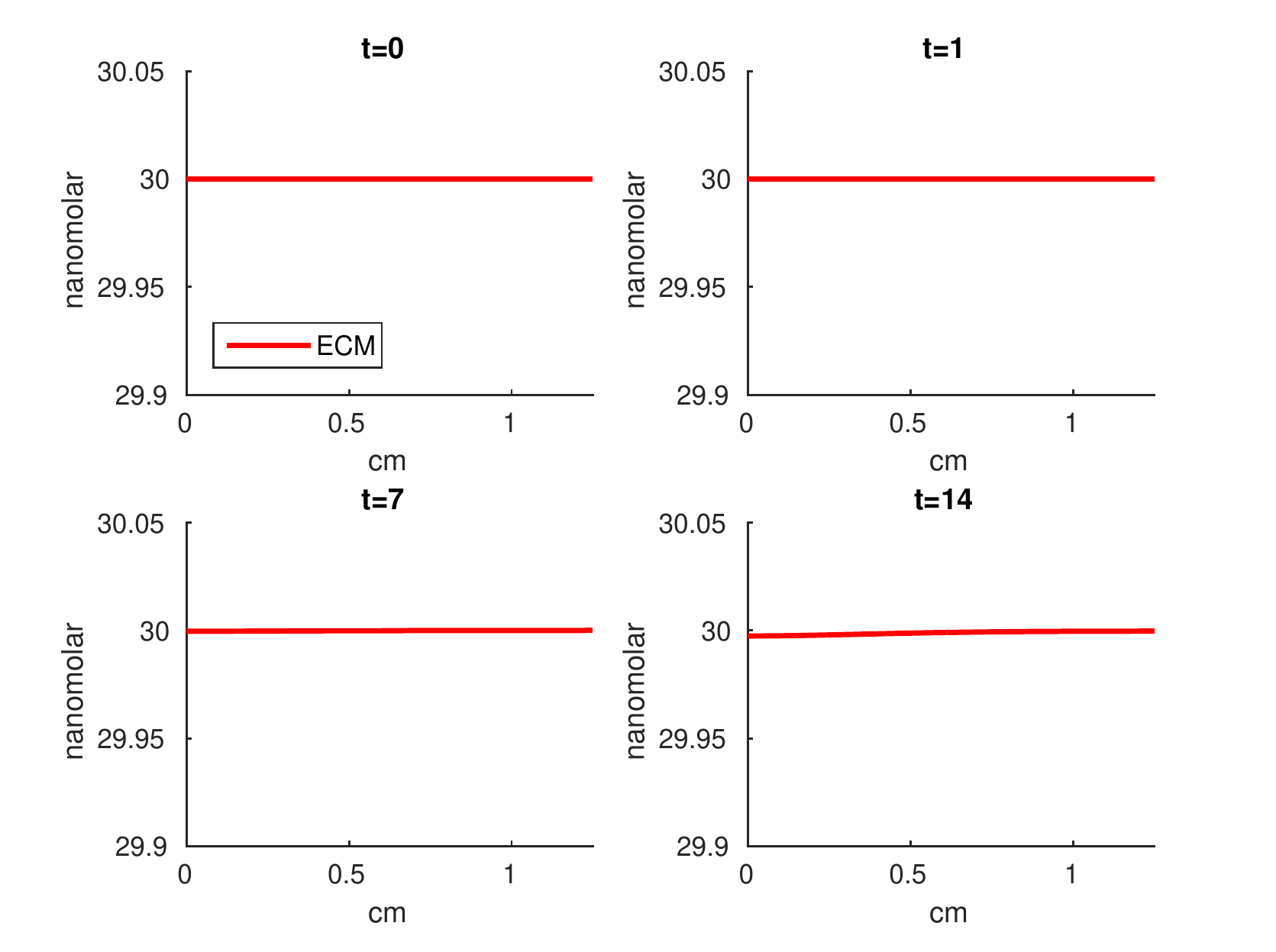}
\caption{Average concentrations of ECM along the radius of the model cylinder at 0, 1, 7, and 14 days. The pressure applied is not enough to produce high deterioration of the ECM.}\label{ECM_ave}  \end{figure}

The results of the simulations with the default parameters are presented in Figures \ref{CU_contour} - \ref{ECM_ave}. Figure \ref{CU_contour} presents contour plots of the density of the $C_U$ cells initially and at days 1, 7, and 14. It shows the decrease in cellular density as time progresses and the $C_U$ cells are signaled to transition. Figures \ref{CU_ave}, \ref{CT_CE_ave}, \ref{ST_SA_ave}, \ref{Chem_ave}, \ref{ROS_ave}, and \ref{ECM_ave} present the average values of the cellular and chemical concentrations at different radial distances. 
Overall, a pressure application of 0.4 MPa does not seem to cause significant damage to the ECM during the first 14 days, as indicated by Figure \ref{ECM_ave}. However, the pressure is still sufficient to trigger the cascade of chemical processes that can lead to the development of OA. In particular, there are small amounts of PIC (evident in Figure \ref{Chem_ave}) and ROS (Figure \ref{ROS_ave}) released and there is a transition of $C_U$ cells into pre-catabolic ($C_T$) and eventually catabolic ($S_T$) cells (Figures \ref{CT_CE_ave} and \ref{ST_SA_ave}). There is a chemical response to the pressure application, but in this particular case, it does not cause disease, at least as indicated at the 14 day mark. The model qualitatively captures the expected behaviors of the chemicals, particularly the PIC and EPO - higher concentrations near the area of contact and diffusion along the cartilage radius, increasing the overall concentrations with time. 

\subsection{Error analysis}
We compared the presented results to three runs using refined discretization: with halved $\Delta a$, with both $\Delta r$ and $\Delta z$ halved, and halving the time step tolerance (essentially halving $\Delta t$). For each comparison, the relative errors were calculated by comparing the values of $C_U$, $S_T$, and EPO at the final time node (14 days) by the formula $rel\ err = max_{(r,z)}abs[(X_b(r,z) - X_c(r,z))]/(X_b(r,z) + 0.0000001)$, where $X_b$ is the value of $C_U$, $S_T$, EPO from the default parameter run at spatial node $(r,z)$ and $X_c$ is the respective value from the comparison run. A value of 0.0000001 was added to avoid division by zero. The $max$ was taken over all spatial nodes. The reported relative error was the maximum of the three relative errors. The maximal relative error for the half $\Delta a$ was 0.0004. The maximal relative error for the halved $\Delta r$ and $\Delta z$ was 0.016 and the maximal relative error for the halved time interval tolerance was 0.0002.

\subsection{Sensitivity analysis}
\begin{table}
\begin{tabular}{|l|l|l|l|}
\hline
Parameter (base value) & Perturbed value & Positive effect& Negative effect\\
\hline
$\beta_{11} (100)$ &50 &- &$S_T$, $S_A$, EPO-,PIC, ROS\\
&75&-&$S_T$, $S_A$\\
&150&$S_T$, $S_A$, EPO, DAMPs, PIC, ROS& -\\
\hline
$\beta_{13}(10)$ &1 & -& $C_E$-, $C_T$-, $S_T$-, $S_A$-, EPO- -,\\
&&&  DAMPs, PIC-, ROS- - \\
&5&-&$C_E$, $C_T$, $S_A$, EPO-, PIC,\\
&&& ROS\\
&15&$C_T$, $C_E$, $S_T$, $S_A$, EPO, PIC, ROS &-\\
&20&$C_T$, $C_E$+, $S_T$, $S_A$+, EPO+, &-\\
&& DAMPs, PIC+, ROS &\\
\hline
$\kappa_1$ (10) &1 &$C_T$, $S_T$+, $S_A$+, EPO, DAMPs,& $C_E$ \\
&&  PIC+, ROS+ &\\
&5&$C_T$, $S_T$, $S_A$&-\\
\hline
$\kappa_2$(10) &1 &$C_T$, $S_T$+, $S_A$+, EPO, DAMPs,& $C_E$\\
&&  PIC+, ROS+ &\\
&5&EPO& $S_A$ \\
&20& $S_A$ &EPO\\
\hline
$\lambda_F$(0.5) &0.1 &$C_T+$, $C_E$, $S_T$+, $S_A$++, EPO+, &ECM\\
&&DAMPs++, PIC+, ROS+ & \\
&0.3&$S_A$, EPO, DAMPs, PIC &-\\
&0.9&-&$S_A$\\
\hline
$\lambda_M$(0.5) &0.1 &$C_E$, $S_T$++, $S_A$+++, EPO++,&$C_T$, ROS-, ECM\\
&&DAMPs++, PIC++& \\
&0.3&$C_T$, $C_E$, $S_T+$, $S_A+$, EPO+, &-\\
&& DAMPs+, PIC+, ROS+ &\\
&0.7&-&$C_E$, $S_T$, $S_A$, EPO-, PIC,\\
&&&  ROS\\
&0.9&-&$C_E$, $S_T$, $S_A$-, EPO- -, \\
&&& DAMPs, PIC, ROS-\\
\hline
$\lambda_R$ (5)&1 &EPO+ &$S_A$, PIC, ROS\\
&3&EPO&-\\
&7&-&EPO\\
&9&-&EPO\\
\hline
$\mu_{DN}$ (0.05)&0.01 &$S_T$, $S_A$, EPO, DAMPs, ROS &-\\
&0.09&-&EPO, DAMPs\\
\hline
\end{tabular}
\caption{Table of parameter perturbation and its effect on numerical outcomes in cellular density and chemical concentration. Positive effect is increasing the compound, negative is decreasing it. A +/- indicates a significant relative increase/decrease of the value. $C_U$ cells were not included in the table. Only significant effects (at least a $\pm 30 \%$ change) of perturbation relative to the base case are included.}\label{Sens_table}
\end{table}
While some of the variables were derived from the experimental literature, fourteen were estimated from previous computational work. Table \ref{Sens_table} presents the parameters and their respective default values and the values within the range over which they caused significant changes in the model's results. When varying a parameter, we set all other parameters at their default values. Table \ref{Sens_table} also presents cell population and chemical concentration results from increasing/decreasing the set parameters, according to relative comparison to the default case. The vast majority of parameter perturbations did not lead to significant qualitative changes in the behavior of the system; the changes did not generally alter the shape of the graphs or the relationship between the values. There were some quantitative differences between runs. Given that many of our parameters are estimated, better approximations can be a topic for future work. However, the model captures, at least qualitatively, expected chemical and cellular behaviors that can result after an injurious pressure load.
\

\begin{figure}
\centering
\includegraphics[width=0.7\textwidth]{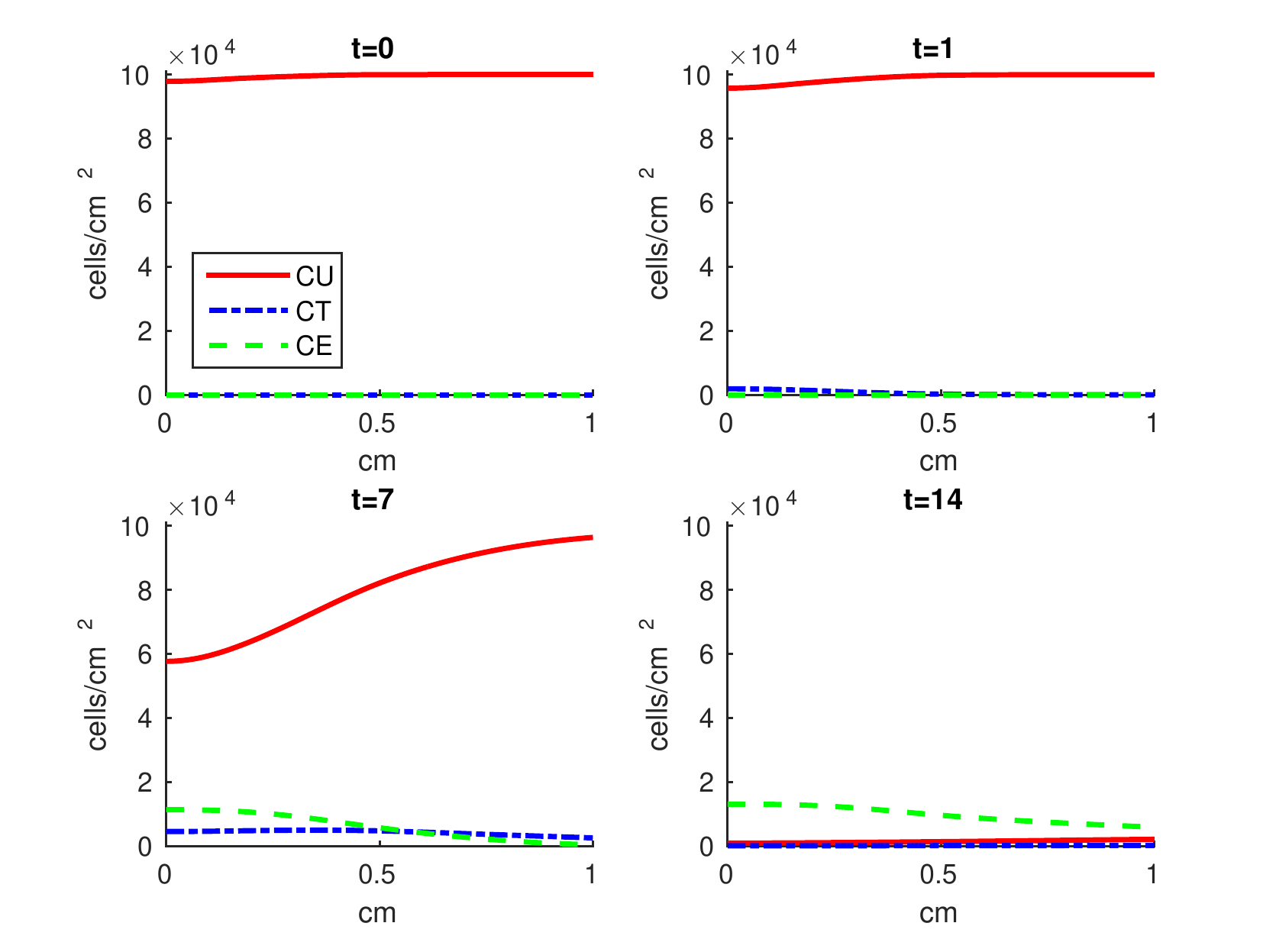}
\caption{The densities of healthy cells ($C_U, C_T,$ and $C_E$) when $\lambda_M = 0.1$ at 0, 1, 7, and 14 days. The densities of $C_U$ cells are close to zero at day 14.} \label{lambda_M_Healthy}  \end{figure}

\begin{figure}
\centering
\includegraphics[width=0.7\textwidth]{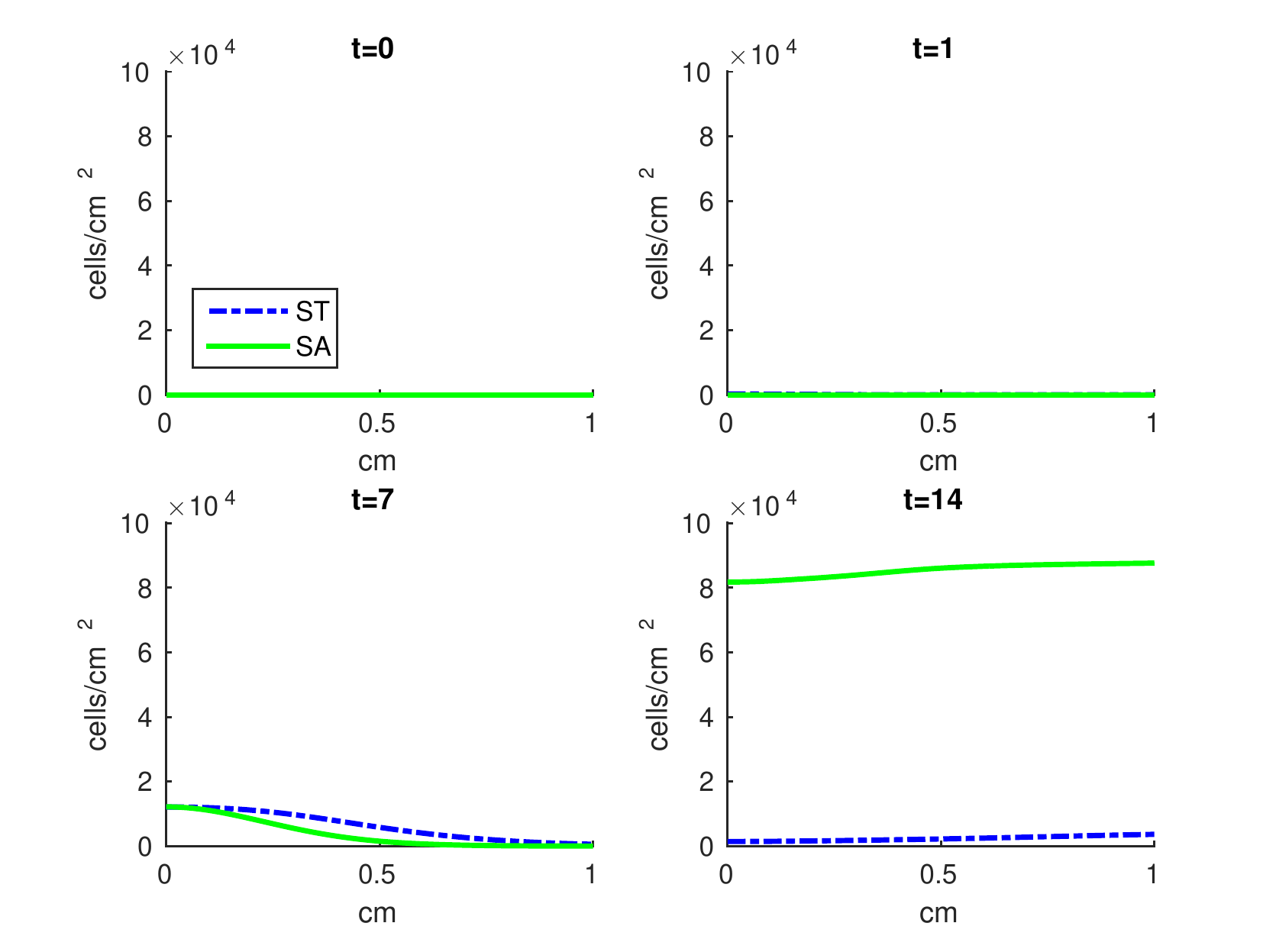}
\caption{The densities of sick cells ($S_T$ and $S_A$) when $\lambda_M = 0.1$ at days 0, 1, 7, and 14. The density of $S_A$ cells is around 80 times as great as in the default parameter case (Figure \protect\ref{ST_SA_ave}).}  \label{lambda_M_Catabolic}\end{figure}

\begin{figure}
\centering
\includegraphics[width=0.7\textwidth]{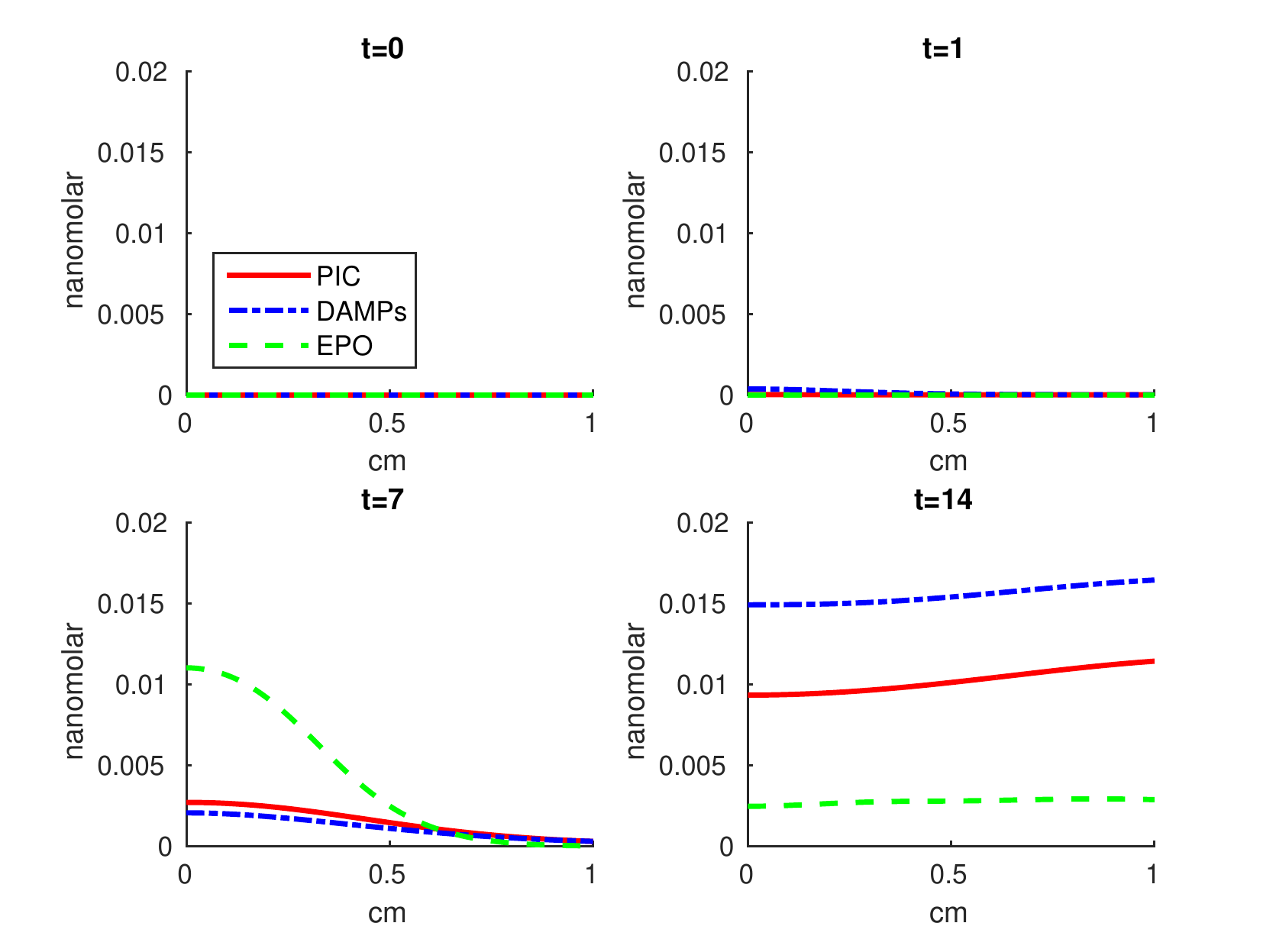}
\caption{The concentrations of EPO, DAMPs, and PIC when $\lambda_M = 0.1$ at days 0, 1, 7, and 14. All chemical concentrations are an order of magnitude higher and more evenly distributed along the radius than in the default parameter case (Figure \protect\ref{Chem_ave}). }\label{lambda_M_Chem}  \end{figure}

\begin{figure}
\centering
\includegraphics[width=0.7\textwidth]{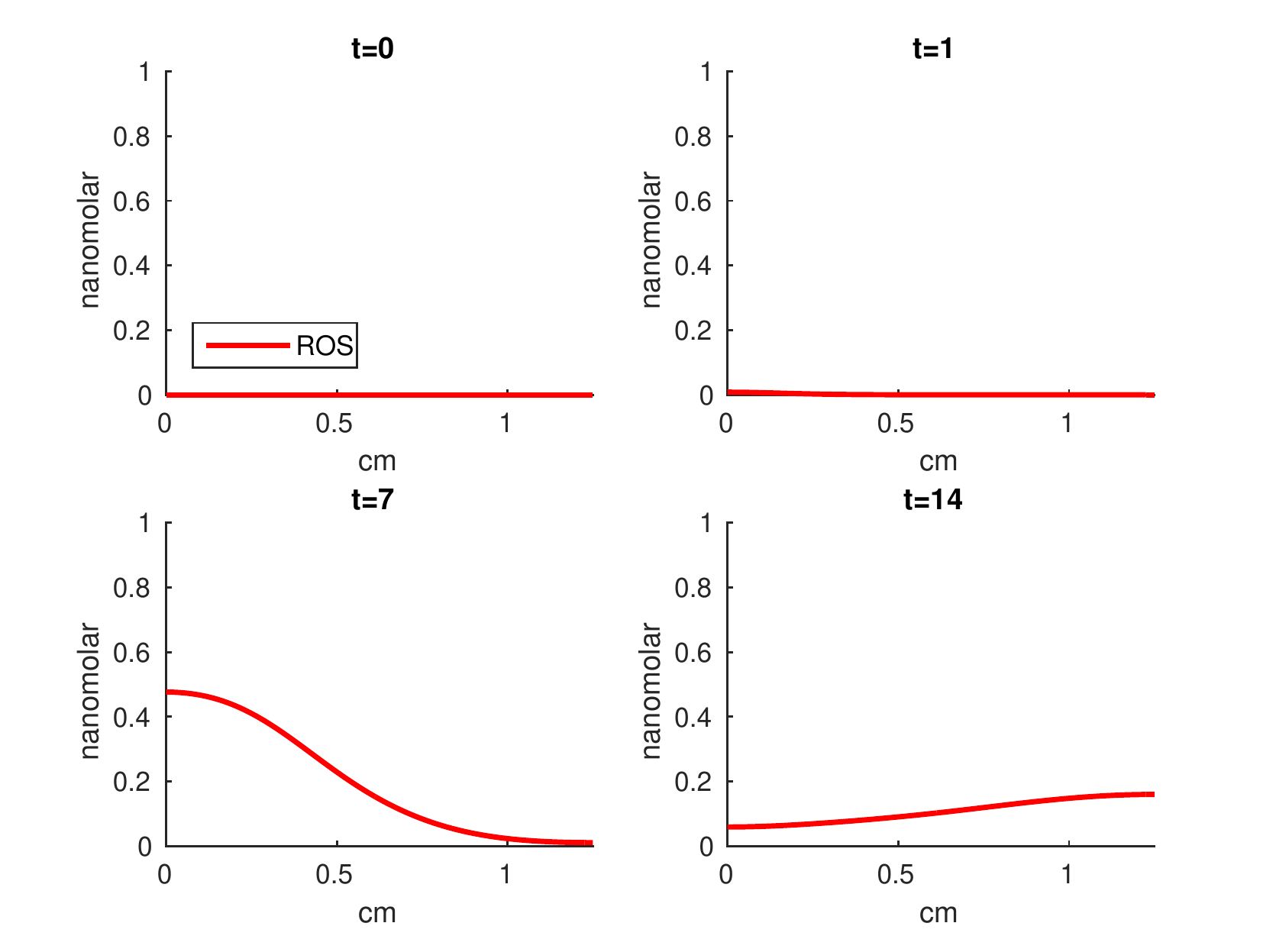}
\caption{The concentrations of ROS when $\lambda_M = 0.1$ at days 0, 1, 7, and 14. The concentration of ROS around the impact site peaks at day 7 and then diffuses significantly by day 14.}  \label{lambda_M_ROS} \end{figure}

\begin{figure}
\centering
\includegraphics[width=0.7\textwidth]{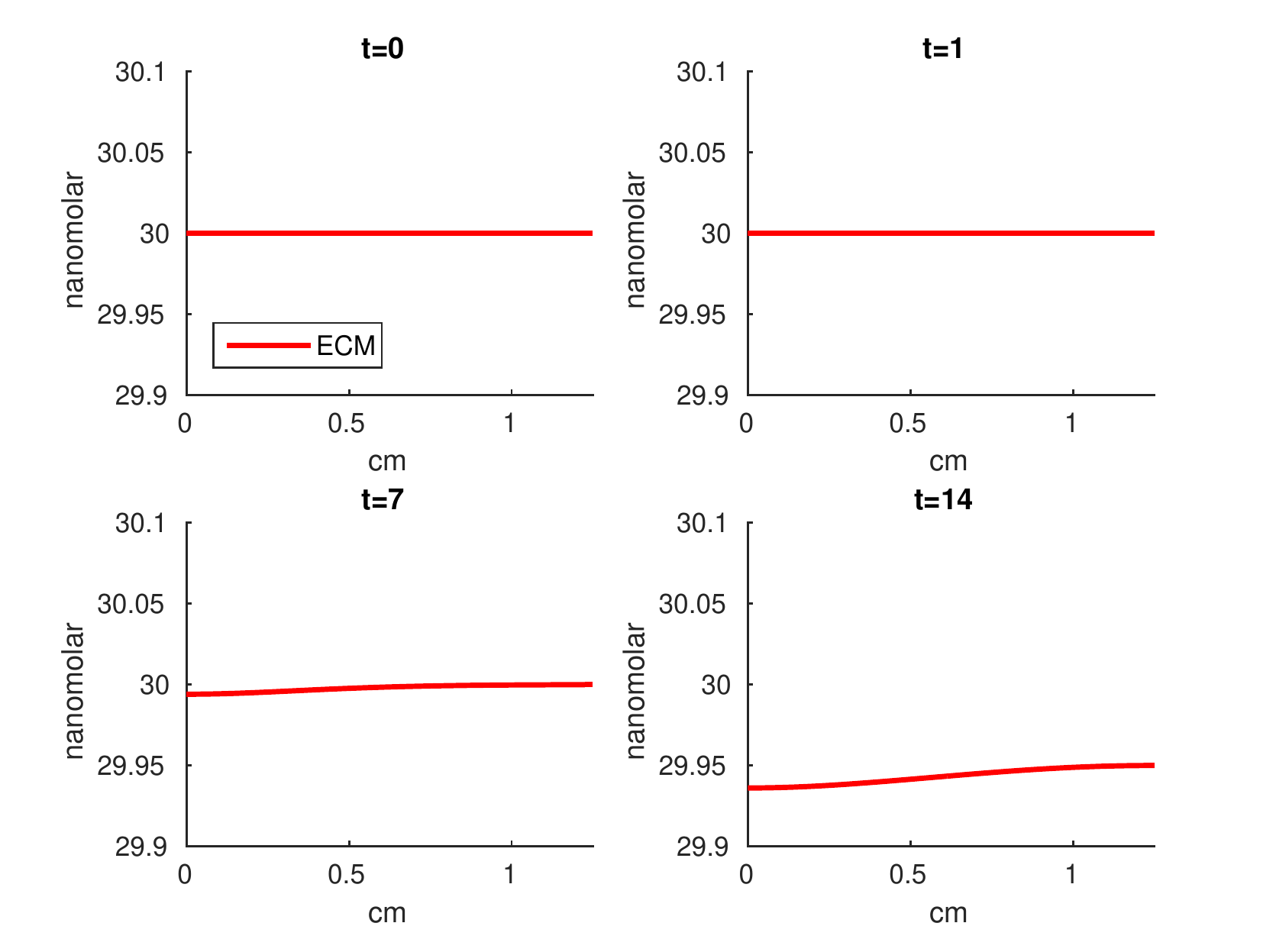}
\caption{The ECM concentrations when $\lambda_M = 0.1$ at days 0, 1, 7, and 14. The ECM has sustained more damage than in the default parameter case (Figure \protect\ref{ECM_ave}).}  \label{lambda_M_ECM} \end{figure}

Notable mentions among the parameters that produced large changes are $\lambda_F$ and $\lambda_M$. Low $\lambda_F$ allows higher saturation of PIC, which leads to more transition of healthy cells into sick cells, as well as higher degradation of the ECM. Therefore, it is not surprising that $S_A$, $S_T$, ROS, and DAMPs all significantly increase when $\lambda_F$ is 0.1. Low $\lambda_M$ allows for higher saturation of DAMPs. Considering that DAMPs released by the necrotic cells after the loading triggers the whole system, the significant effect of $\lambda_M$ perturbations on the system is expected. Some differences that resulted from low $\lambda_M$ are the near eradication of $C_U$ cells and a more pronounced diffusion of the chemicals by day 14. To see the significant changes in behavior in the system when $\lambda_M$ is 0.1, refer to Figures \ref{lambda_M_Healthy} - \ref{lambda_M_ECM}.

\section{Discussion}\label{Discussion_section}
We presented a multiscale mathematical model of the balancing act between the pro- and anti- inflammatory cytokines released by the chondrocytes in articular cartilage under the application of pressure from an indenter onto a cylindrical cartilage explant. Our current model differs from previous work by incorporating a mechanical finite element simulation and analysis to estimate the role of the initial loading on the biochemical components of the model. This addition also requires an extra spatial dimension for the cartilage (depth). The model includes three components, chemical, cellular, and mechanical (tissue-scale), instead of the two components of the previous work. The mechanical component describes the strain and cell death resulting from the initial pressure application. The cellular component describes the different states of the chondrocytes (healthy, sick, dead), and the different chemicals being in these states makes them release. The chemical component represents the interplay between chemicals, their signaling to cells to switch states, and their release by the chondrocytes in a particular state. The mechanical component was modeled using linear elasticity, solved with the finite element solver Abaqus\texttrademark . The chemical component was modeled by reaction-diffusion equations, and the cellular component was modeled by age-structure, to capture the cellular age of state-switching.
\

The numerical results of the model simulated the anticipated chemical and cellular behaviors well. Concentrations of chemicals and different cellular states are highest around the center of the cylinder, which was expected, given that this is the position of the pressure application. Therefore, the model is successful in qualitatively estimating the effect of injurious pressure application on the cartilage explant for 14 days. For a better quantitative estimates we would need a better parametrization, a better understanding of the sensitivity of the system to perturbations of the unknown parameters and the initial strains, and experimental validation.  Longitudinal data over a longer time frame would further allow us to validate and calibrate our model to capture the development of OA in the long run.
\

 The sensitivity analysis provided no surprises; the model's responses to perturbations were all biologically reasonable and expected given the equations and their dependence on the perturbed parameters. The only parameter change that produced a vastly different system behavior was setting $\lambda_M$ to 0.1 (see Figures \ref{lambda_M_Healthy} - \ref{lambda_M_ECM}).
 \

One limitation of the model is that it does not include cartilage heterogeneity. Making the model seem more mechanistic by using biphasic elasticity would introduce additional modeling and computational complexity. In the end, we may not have a significantly more accurate model. Furthermore, linear elasticity seems to be a sufficient approximation for the behavior of cartilage under many physiological loading conditions \cite{Carter:1999}. A possible inclusion of more complicated mechanical properties can be a topic for future work.
\

The main goal of the presented model was the integration of the explicit mechanics, using finite element analysis, into a biomathematical model. The interaction between the finite element simulations and the mathematical modeling of biochemical processes presents an important framework for collaboration between biomedical engineering and mathematical modeling and simulation, and can lead to fruitful collaborations resulting in understanding, preventing, and treating OA.

\section*{Acknowledgments}
All authors were partially supported by the NIAMS: CORT Innovations to Assess and Forestall Post-Traumatic Osteoarthritis 5 P50 AR055533-05. The authors would like to thank Prof.~Douglas Pedersen for allowing us access to his Abaqus\texttrademark license and for helpful comments and ideas during discussions of the manuscript.

\section*{Conflict of interest statement}
The authors declare that there are no conflicts of interest.

\newpage


\end{document}